# Schmidt information and entanglement in quantum systems

A.Yu. Bogdanov**, Yu.I. Bogdanov*, K.A. Valiev*

*Institute of Physics and Technology, Russian Academy of Sciences*
**Faculty of Physics, Moscow M.V. Lomonosov State University, Russia*

The purpose of this paper is to study entanglement of quantum states by means of Schmidt decomposition. The notion of Schmidt information which characterizes the non-randomness of correlations between two observers that conduct measurements on EPR-states is proposed. In two important particular cases – a finite number of Schmidt modes with equal probabilities and Gaussian correlations - Schmidt information is equal to Shannon information. A universal measure of a dependence of two variables is proposed. It is based on Schmidt number and it generalizes the classical Pearson correlation coefficient. It is demonstrated that the analytical model obtained can be applied to testing the numerical algorithm of Schmidt modes extraction. The introduced notions of information and correlations based on Schmidt decomposition naturally generalize the corresponding classical notions. A thermodynamic interpretation of Schmidt information is given. It describes the level of entanglement and correlations of a micro-system with its environment.

**Introduction**

Quantum informatics has been intensively studied recently. That is due to the fact that quantum computers and quantum cryptography devices can dramatically increase the efficiency of solving many important problems compared to classical computers [1]. One of the main notions of quantum informatics is entanglement. It was first introduced and analyzed in the famous work by Einstein, Podolsky and Rosen in 1935 in the form of a so-called EPR- paradox [2]. Imagine that we have two particles that interacted in the past. Whatever time passes, the particles continue to stay in the entangled state that is characterized by specific quantum correlations. For instance, if one conducts measurements on a particle then one gets the information about the other particle. Furthermore, the particles can be far from one another. That phenomenon puts a question on quantum mechanics' completeness and locality. Later, Bohr demonstrated that the description of EPR-pairs does not contain any paradoxes or inconsistencies by introducing his correspondence principle [3]. Still, due to the uncommonness of quantum properties, the question remained a theoretical one and it mainly concerned the philosophy and methodology of quantum mechanics' interpretation. The situation significantly changed after the well-known works by Feynman that initiated a new field in science namely quantum informatics [4]. From that moment the questions concerning quantum states' entanglement moved from the theoretical field to the practical one.

It is quite remarkable that mathematical tools for describing the entanglement were proposed as far as in 1906 by Schmidt [5].  (For basic notions of the theory see [1, 6])

At first, in scientific works the entanglement of discrete degrees of freedom connected with spin and polarization of particles was considered. Recently, the entanglement of continuous degrees of freedom is widely discussed (related to coordinate, momentum, frequency, etc.) [7, 8, 9].

In the present work we consider informational aspects that are connected with the Schmidt decomposition and the entanglement of quantum states.

In part 1 we introduce the notion of Schmidt information as a measure of non-randomness of correlations between two observers that conduct measures on EPR quantum states.

In part 2 we use the mathematical apparatus of Schmidt decomposition and the notion of entanglement to construct an analytical model that allows one to model Gaussian correlations of classical probability theory. It appears that in the analytical model Schmidt information equals to Shannon information.

---

* E-mail: bogdan@ftian.oivta.ru

The relation between the Pearson correlation coefficient and the Schmidt number was discovered. It allows one to construct a universal measure of dependence between two variables. In the considered approach the criteria of the dependence and the entanglement of two sub-systems that form a unified physical system are generally the same notions.

In part 3 the considered analytical model serves for testing the numerical algorithm of Schmidt modes extraction that was proposed in [10]. It is demonstrated that the results of analytical and numerical calculations are almost the same.

In part 4 we provide a thermodynamic interpretation of the introduced notions for an example of a harmonic oscillator in a thermostat.

Finally, in part 5 we formulate the principle results of the work.

**1. Information measure based on Schmidt decomposition**

Let the probability amplitude (wave function) $\psi(p,q)$ of system be a function of two variables $p$ and $q$, where both variables can be either one-dimensional or multi-dimensional. Schmidt decomposition has the following form [1,5,6]:

$$\psi(p,q) = \sum_k \sqrt{\lambda_k}\, \psi_k^{(1)}(p)\psi_k^{(2)}(q), \qquad (1)$$

where $\lambda_k$ are the weight multipliers that satisfy the normalization condition

$$\sum_k \lambda_k = 1 \qquad (2)$$

We assume that the summands in the decomposition (1) are presented in the order of non-increasing $\lambda_k$

In numerical calculations the considered function $\psi(p,q)$ is presented in a discrete form by a matrix $\psi_{j_1 j_2} = \psi(p_{j_1}, q_{j_2})$ where $1 \leq j_1 \leq n$, $1 \leq j_2 \leq n$. Let the function is defined on a uniform square mapping of size $n \times n$.

The number of discretization points is to be quite large (see part 3). Note that in the numerical algorithm one may use rectangular mappings, where the number of discretization points is different for each variable.

$\psi_k^{(1)}(p)$ and $\psi_k^{(2)}(q)$ are called Schmidt modes. The number of weight multipliers $\lambda_k$ in the decomposition (1) and the corresponding number of Schmidt modes can be both finite and infinite.

The principal numerical characteristic of Schmidt decomposition is the Schmidt number $K$ that characterizes the effective number of modes in Schmidt decomposition.

$$K = \frac{1}{\sum_k \lambda_k^2} \qquad (3)$$

Due to the definition, according to the normalization condition for $\lambda_k$, the number $K$ is not less than unity. It is equal to unity only if there is a single non-zero summand in Schmidt decomposition.

Schmidt decomposition provides a vivid mathematical apparatus for entanglement visualization and analysis. For instance, if the first observer registers the variable $p$ in state $\psi_k^{(1)}(p)$ then the variable $q$ is to be registered in state $\psi_k^{(2)}(q)$ (for the same $k$)

The spatially divided observers are usually named Alice and Bob in quantum cryptography. Let symbol $a_k$ correspond to mode $k$. Then a set of symbols $a_1, a_2, \ldots$ is either finite or infinite. By registering EPR-states in a basis defined by Schmidt modes, Alice and Bob register some string of numbers $k_1, k_2, \ldots$ and, therefore, some string of symbols $a_{k_1}, a_{k_2}, \ldots$. If we do not take into account mistakes and noise then we may claim that Alice and Bob observe the same string (100% correlation) where symbol $a_k$, $k = 1, 2, \ldots$ has the probability $\lambda_k$, $k = 1, 2, \ldots$.

Let us provide a vivid illustration for the Schmidt number. Let the length of the completely correlated strings that Alice and Bob possess be equal to $n$ ($n$ representatives of a quantum EPR-state were measured.) For any finite $n$ we have a non-zero probability that the coincidence of Alice's and Bob's strings is accidental. In other words we may assume that the strings are generated not by one EPR – source but by two independent sources where each arbitrary symbol $a_k$, $k = 1, 2, \ldots$ has the probability $\lambda_k$, $k = 1, 2, \ldots$ defined above.

For a separate probability experiment on two independent sources we may obtain an arbitrary combination of symbols $a_{k_1} a_{k_2}$; $k_1, k_2 = 1, 2, \ldots$. Due to the normalization condition (2) we may write $\sum_{k_1, k_2} \lambda_{k_1} \lambda_{k_2} = 1$. In the latter sum we are only interested in the summands for which $k_1 = k_2 = k$. Thus, the probability of an accidental coincidence of symbols is given by

$$P^{(1)} = \sum_k \lambda_k^2 = \frac{1}{K} \qquad (4)$$

The probability of an accidental coincidence in a series of $n$ independent experiments is given by

$$P^{(n)} = \left(\frac{1}{K}\right)^n = 2^{-n \log_2(K)} = \exp(-n \ln(K)) \qquad (5)$$

According to the equation (5) let us introduce a new notion of information based on Schmidt number. The corresponding information characterizes the measure of non-randomness of the correlation between two considered observers. The information that is contained in a string of $n$ letters of a statistical ensemble is

$$I_2 = n \log_2(K) \qquad (6)$$
$$I_e = n \ln(K) \qquad (7)$$

Here $I_2$ and $I_e$ is the information measured in binary and natural units correspondingly.

The introduced measure of information is additive (proportional to the sample size).

It can be expressed in the form similar to the Boltzmann equation

$$I_e = \ln(W), \text{ where } W = K^n = \exp(n \ln(K)) \qquad (8)$$

According to (5) and (8), the probability of an accidental coincidence is equal to $\frac{1}{W}$. The parameter $W$ can be interpreted as an effective number of equally probable microstates. The number $W$ does not need to be whole. It can be interpreted by an urn scheme. Imagine that there are $W$ volumes of sand. Only one of the volumes is dark, whereas other volumes are bright. Then the sand is mixed and one grain of sand is taken. The probability for it to be dark is $\frac{1}{W}$. Thus we move from classical probabilities to geometrical probabilities.

The phenomenon of entanglement can be interpreted using the notion of entropy. The entropy of entanglement is defined as follows: (e.g. see [1,10])

$$S = -\sum_k \left(\lambda_k \log \lambda_k\right) \qquad (9)$$

As the base of the logarithm one may choose $2$, $e$ or some other number.

Before the measurement the state is non-factorized (i.e. entangled) and it contains indefiniteness concerning future measurements (the entropy of entanglement is positive). After a measurement the state is factorized, there is no indefiniteness and the Schmidt number is equal to unity (the entropy of entanglement is equal to zero).

Thus, the equations (6)-(9) provide a measure of decreasing indefiniteness concerning future measurements, i.e. the measure of the information obtained as a result of a measurement and the corresponding break-up of the initial EPR-representatives.

In two important particular cases the information based on Schmidt decomposition and Shannon information coincide. The first case is a finite number of Schmidt modes with equal probabilities (each mode has the probability $\frac{1}{K}$). The second special case concerning Gaussian correlations is described in the next section.

**2. Gaussian correlations. Schmidt correlation coefficient**

A two-dimensional normal distribution that defines a common distribution of random values $x_1$ and $x_2$ with a correlation coefficient $\rho$ is defined by the following probability density [11]:

$$p(x_1, x_2) = \frac{1}{2\pi\sigma_1\sigma_2\sqrt{1-\rho^2}} \exp\left(-\frac{1}{2(1-\rho^2)}\left[\frac{(x_1-m_1)^2}{\sigma_1^2} - \frac{2\rho(x_1-m_1)(x_2-m_2)}{\sigma_1\sigma_2} + \frac{(x_2-m_2)^2}{\sigma_2^2}\right]\right) \qquad (10)$$

Here $m_1, m_2, \sigma_1^2, \sigma_2^2$ are the corresponding mean values and variances of the random variables.

Let us express the distribution as a quantum state realization with the following wave function:

$$\psi(x_1, x_2) = \sqrt{p(x_1, x_2)} \qquad (11)$$

Let us note the crucial difference between classical distributions and quantum states. When we make a transition from a distribution density to a wave function, we obtain a much richer quantum state. In principle, quantum state (11) includes not only the considered distribution (10) that arises when we conduct measurements in a coordinate representation, but also all the information about other (complementary) distributions, e.g. momentum distribution. According to Bohr's complementarity principle, for a complete description of all possible experiments related to quantum states one needs to derive a set of mutually-complementary distributions, but not only one statistical distribution [12].

Let us apply Schmidt decomposition to the wave function. As a result, we get a set of Schmidt modes that can be expressed by Chebyshev-Hermit polynomials ($C_k^{(1)}, C_k^{(2)}$ - are normalization constants, $k = 0, 1, \ldots$):

$$\psi_k^{(1)}(x_1) = C_k^{(1)} H_k\left(\frac{(x_1 - m_1)}{\sigma_1}\sqrt{\frac{K}{2}}\right)\exp\left(-K\frac{(x_1 - m_1)^2}{4\sigma_1^2}\right) \quad (12)$$

$$\psi_k^{(2)}(x_2) = C_k^{(2)} H_k\left(\frac{(x_2 - m_2)}{\sigma_2}\sqrt{\frac{K}{2}}\right)\exp\left(-K\frac{(x_2 - m_2)^2}{4\sigma_2^2}\right) \quad (13)$$

It appears that the Schmidt number is related to the correlation coefficient as:

$$K = \frac{1}{\sqrt{1-\rho^2}} \quad (14)$$

The weights in Schmidt decomposition set a geometric progression with the following parameters:

$$\lambda_0 = \frac{2}{K+1} \quad (15)$$

which is the weight of the principal (zero) Schmidt component

$$q = \frac{K-1}{K+1} \quad (16)$$

which is the multiplier of the geometric progression

Shannon information (mutual information contained in the joint distribution of the random variables $x_1$ and $x_2$) is defined by the following equation:

$$I_{Sh}(x_1; x_2) = \int p(x_1, x_2) \log\left(\frac{p(x_1, x_2)}{p_1(x_1) p_2(x_2)}\right) dx_1 dx_2 \quad (17)$$

Here $p(x_1, x_2)$ is the density of the joint distribution.

The densities $p_1(x_1)$ and $p_2(x_2)$ define marginal distributions of random variables $x_1$ and $x_2$ correspondingly.
They are defined as following:

$$p_1(x_1) = \int p(x_1, x_2) dx_2, \quad p_2(x_2) = \int p(x_1, x_2) dx_1 \quad (18)$$

A calculation of Shannon information for a two-dimensional normal distribution leads to the following important result

$$I_{Sh}(x_1; x_2) = \log(K) \quad (19)$$

Thus, for Gaussian correlating variables Shannon information has the same value as Schmidt information. The information based on Schmidt decomposition is a measure of non-randomness of EPR-type correlations. At the same time, the traditional Shannon mutual information characterizes the measure of data compression that is possible due to the mutual dependence of the considered variables.

Using the definition (9) it is possible to demonstrate that in that case the entropy of entanglement has the form:

$$S = \log\left(\frac{K+1}{2}\right) + \frac{(K-1)}{2}\log\left(\frac{K+1}{K-1}\right) \qquad (20)$$

Let us rewrite equation (14) that expresses the interrelation between the correlation coefficient and the Schmidt number in the following form:

$$\rho^2 = 1 - \frac{1}{K^2} \qquad (21)$$

That relation is valid for Gaussian distribution and it can be set as a definition of the correlation coefficient between two arbitrary random values (and with EPR-state variables).

The measure of correlation defined according to (21) has an evident advantage compared to the conventional correlation coefficient (Pearson). From (21) one can easily see that $0 \leq \rho^2 \leq 1$, while the relation $\rho = 0$ corresponds to independent values of $x_1$ and $x_2$. From the other hand, it is known [11] that for the traditional definition of a correlation coefficient from the independence follows non-correlation whereas the reverse is not true. Variables do not have to correlate; still they can be dependent because the traditional correlation coefficient may serve only as a measure of linear dependence between variables. For the new definition of the correlation measure the terms correlation and dependence are synonyms.

## 3. A comparison of the results of analytical and numerical calculations

Most problems related to a study of entanglement in quantum informational systems can not be treated analytically. A comparison of numerical calculations results with the precise solution presented above reveals that one may obtain the highest precision (a coincidence up to 15-16 digits). Such a precision is constrained by an algebraic calculation error. An example of such a comparison is given in Fig. 1 and Table 1.

Table 1. Test of the numerical algorithm dependent on the number of discretization points

| $\lambda_i$ | Theory | $n \times n = 30 \times 30$ | $n \times n = 50 \times 50$ | $n \times n = 100 \times 100$ |
|---|---|---|---|---|
| 1 | 0.607135541614981 | 0.60647628536655 | 0.607135554628264 | 0.607135541614983 |
| 2 | 0.238521975722865 | 0.24466551241268 | 0.23852190856412 | 0.238521975722865 |
| 3 | 0.0937068068052879 | 0.08744563654242 | 0.09370729417429 | 0.0937068068052879 |
| 4 | 0.03681407390254 | 0.04862772916714 | 0.03681200534047 | 0.0368140739025491 |
| 5 | 0.014462941204671 | 0.00806011821403 | 0.01446967071109 | 0.0144629412046711 |
| 6 | 0.0056819755630274 | 0.00439073338559 | 0.00566756454034 | 0.0056819755630275 |
| $K$ | 2.29415733870562 | 2.2843024512965 | 2.29415712995952 | 2.29415733870561 |

We consider an example that corresponds to the following parameters: $\rho = 0.9$, $m_1 = 1$, $m_2 = -1$, $\sigma_1 = 2$, $\sigma_2 = 1$

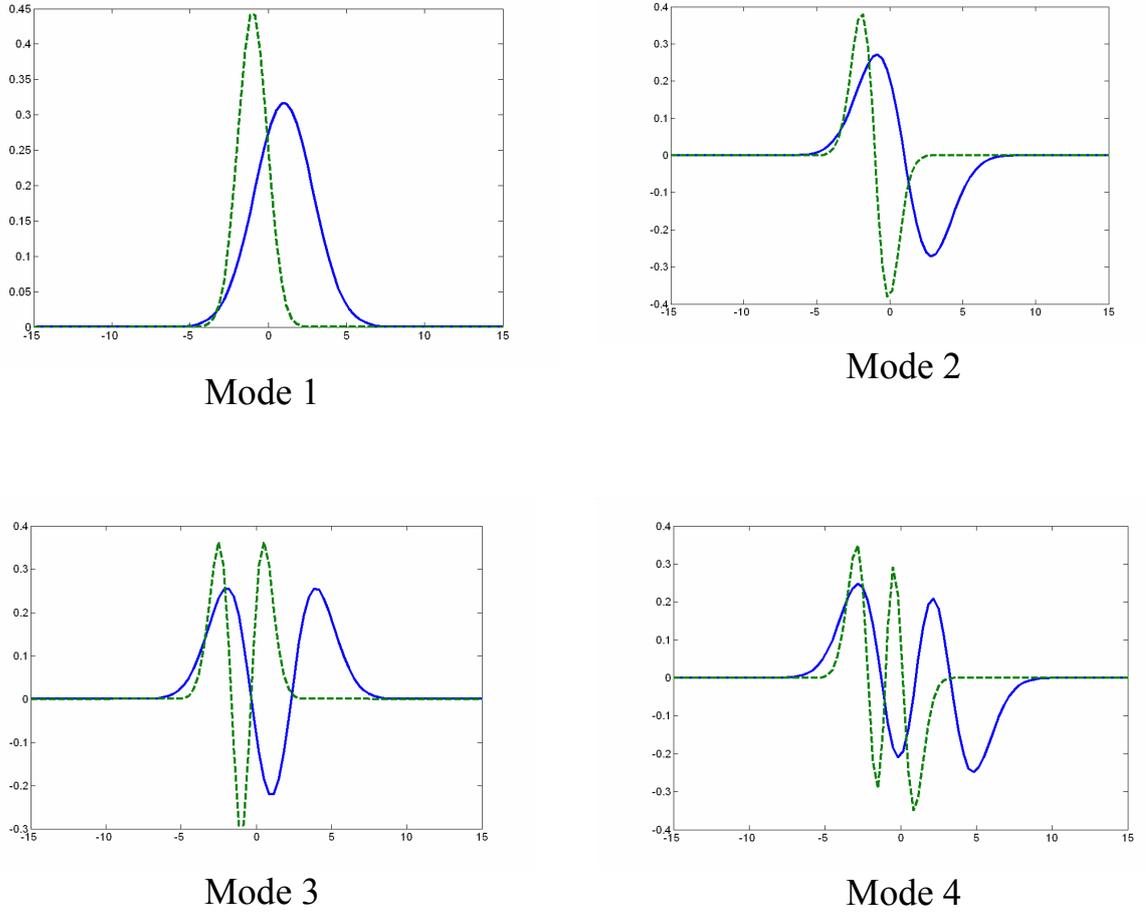

Figure 1 An illustration of Schmidt modes

On fig. 1 the primary four Schmidt mode pairs are presented that correspond to the considered example.

In table 1 the results of calculations of the primary six weight coefficients of Schmidt decomposition and Schmidt numbers are presented. Already, an almost coincidence is observed for grid $n \times n = 100 \times 100$.

**4. A thermodynamic interpretation**

The modes (12)-(13) have the same form as harmonic oscillator states. From the other hand, the probabilities of harmonic oscillator states make a geometric progression due to the spectrum equidistance. By analogy, the weight coefficients that define a probability of the corresponding Schmidt modes registration also make a geometric progression as shown above. Let us set a correspondence between the coefficients of the geometric progression (16) and a Boltzmann multiplier that corresponds to some temperature $\theta$.

$$\exp(-\frac{\hbar\omega}{\theta}) = \frac{K-1}{K+1} \qquad (22),$$

Where $\omega$ is the oscillator frequency, $\hbar$ is the Planck constant.

From (22) we get the following relation between the temperature and the Schmidt number that is defined by hyperbolic cotangent:

$$K = \coth\left(\frac{\hbar\omega}{2\theta}\right) \qquad (23)$$

According to (21) and (23) the relation between the temperature and the correlation coefficient is given by an equation that includes a hyperbolic cosine:

$$\rho^2 = \frac{1}{\cosh^2\left(\frac{\hbar\omega}{2\theta}\right)} \qquad (24)$$

We may suppose that coordinate $x_1$ defines the coordinate of a micro-system while the variable $x_2$ corresponds to the macroscopic environment (thermostat)

According to (23) and (24) for low temperatures $\theta << \frac{\hbar\omega}{2}$ there is no entanglement between a quantum system and the environment ($K \to 1$). Then the micro-system and the environment do not correlate with each other ($\rho^2 \to 0$).

On the reverse, for high temperatures $\theta >> \frac{\hbar\omega}{2}$ a quantum system and the environment are highly entangled ($K \to \infty$). Then they are highly correlated ($\rho^2 \to 1$).

Using (22)-(23) one may show that the entropy of entanglement (20) coincides with the well known quantum statistical notion of entropy that corresponds to oscillator degrees of freedom [13]:

$$S = -\log\left(1 - \exp\left(-\frac{\hbar\omega}{\theta}\right)\right) + \frac{\hbar\omega}{\theta}\frac{1}{\left(\exp\left(\frac{\hbar\omega}{\theta}\right) - 1\right)} \qquad (25)$$

Note that the idea that a thermodynamic equilibrium is a result of an entanglement between a system and its environment has been widely discussed recently [14].

The considered example demonstrates a situation when the initial pure state that describes the entanglement of two subsystems yields a distribution (after the measurement) that precisely corresponds to a heat equilibrium state of one of the subsystems in the thermostat.

**5. Conclusion**

Let us formulate the principal results of the work.
1. The notion of Schmidt information as a measure of non-randomness between two observers conducting measurements upon EPR quantum states is defined. For a finite number of Schmidt modes with uniform distribution and for Gaussian correlations Schmidt information and Shannon information coincide.
2. An analytical model that is based on a use of an entangled quantum state is proposed. It allows one to model Gaussian correlations of the classical probability theory. An interrelation between Pearson correlation coefficient and Schmidt number is demonstrated. A universal measure for describing the dependence between variables based on the notion of entanglement is proposed.

3. It is demonstrated that the proposed analytical model may be applied for testing numerical algorithms of quantum state entanglement investigation.
4. Based on the example of a quantum oscillator in a thermostat it is demonstrated that Schmidt information can effectively describe the level of entanglement and of correlations of a micro-system with its environment.